\documentclass[
 amsmath,amssymb,aps,twocolumn,a4paper,
]{revtex4-1}

\usepackage{graphicx}
\usepackage{dcolumn}
\usepackage{bm}
\usepackage{lipsum}

\newcommand{\mv}[1]{\ensuremath{\mathbf{#1}}} 
\newcommand{\gv}[1]{\ensuremath{\mbox{\boldmath$ #1 $}}} 
\newcommand{\uv}[1]{\ensuremath{\mathbf{\hat{#1}}}} 
\newcommand{\abs}[1]{\left| #1 \right|} 
\newcommand{\avg}[1]{\left \langle #1 \right \rangle} 

\usepackage[colorlinks]{hyperref}  
\usepackage{cleveref}
\hypersetup{linkcolor=blue,citecolor=blue,urlcolor=blue}

\usepackage{titlesec}
\titleformat{\section}
  {\centering\normalfont\fontsize{11}{15}\bfseries}{\thesection}{1em}{}

\usepackage[toc]{appendix}

\crefname{figure}{fig.}{figs.}
\Crefname{figure}{Figures}{Figures}

\begin{document}

\setcounter{page}{1} 

\title{Mode-coupling theory for mixtures of athermal self-propelled particles}

\author{Vincent E. Debets$^{1,2}$, Liesbeth M.C. Janssen$^{1,2*}$}

\affiliation{$^{1}$Department of Applied Physics, Eindhoven University of Technology, P.O. Box 513,
5600 MB Eindhoven, The Netherlands\\$^{2}$Institute for Complex
Molecular Systems, Eindhoven University of Technology, P.O. Box 513,
5600 MB Eindhoven, The Netherlands\\}
\email{l.m.c.janssen@tue.nl}

\begin{abstract}%
{\noindent Dense or glassy active matter, as a result of its remarkable resemblance to passive glass-forming materials, is enjoying increasing scientific interest. To better grasp the subtle effect of active motion on the process of vitrification, a number of active mode-coupling theories (MCTs) have recently been developed. These have proven capable of qualitatively predicting important parts of the active glassy phenomenology. However, most efforts so far have only considered single-component materials, and their derivations are arguably more complex than the standard MCT case, which might hinder broader usage. Here we present a detailed derivation of a distinct active MCT for mixtures of athermal self-propelled particles that is more transparent than previously introduced versions. The key insight is that we can follow a similar strategy for our overdamped active system as is typically used for passive underdamped MCT. 
Interestingly, when only considering one particle species, our theory gives the \emph{exact} same result as the one obtained in previous work which employed a highly different mode-coupling strategy.
Moreover, we assess the quality of the theory and its novel extension to multi-component materials by using it to predict the dynamics of a Kob-Andersen mixture of athermal active Brownian quasi-hard spheres. We demonstrate that our theory is able to capture all qualitative features, most notably the location of the optimum of the dynamics when the persistence length and cage length coincide, for each combination of particle types.}
\end{abstract}

\maketitle 

\section*{Introduction}
\noindent Motivated by its frequent occurrence in the biological realm, active matter has enjoyed much scientific interest during the past two decades~\cite{Bechinger2016,Ramaswamy2010,Marchetti2013rev} and is now firmly positioned at the forefront of soft matter and biophysical research. Initial active matter studies, e.g., to explain the swimming behavior of microorganisms, have been predominantly geared towards self-propelled particles in the now reasonably well-understood dilute limit~\cite{Bechinger2016}. By contrast, the high-density regime has 
only recently seen a significant upsurge of interest~\cite{Janssen2019active,Berthier2019review}. 
Dense active matter has already been shown to exhibit a myriad of exciting non-equilibrium phenomena, such as velocity ordering~\cite{Caprini2020velocity1,Caprini2020velocity2}, motility induced phase separation (MIPS)~\cite{Digregorio2018,Geyer2019,Caporusso2020,Omar2021,paoluzzi2021,Cates2015}, collective actuation~\cite{Baconnier2022}, and activity-induced crystallization~\cite{Briand2016,Ni2014}. 
So-called active glassy materials constitute an arguably even more extreme example of dense active matter. 
These are disordered assemblies of self-propelled particles that, irrespective of their intrinsic driving, have undergone kinetic arrest. Experimental examples can be found in the context of cell layers~\cite{Angelini2011cell,Garcia2015cell,Grosser2021,Lama2022}, individual cells~\cite{Zhou2009cell,Parry2014bacterial,Nishizawa2017cell}, granular matter~\cite{Arora2022}, and colloids~\cite{Klongvessa2019colloid1,Klongvessa2019colloid2}, and they have also been witnessed in many theoretical and simulation studies~\cite{Voigtmann2017,SzamelABP2019,SzamelAOUP2015,SzamelAOUP2016,FengHou2017,BerthierABP2014,DijkstraABP2013,BerthierAOUP2017,Berthier2013activeglass,Flenner2020,FlennerAOUP2016,Henkes2011active,Reichert2020modecoupling,Reichert2020tracer,Reichert2021rev,Nandi2018,Sollich2020,janzen2021aging,Janssen2017aging,Bi2016cell,paoluzzi2021,Debets2021cage,Keta2022,Debets2023chiral}. As a result of the increasingly strong role of particle-particle interactions at high densities, active glassy materials show a remarkable resemblance to their passive counterparts~\cite{Janssen2019active,Debenedetti2001supercooled,binder2011}. However, this does not imply that activity is completely overshadowed by interactions as it can indeed influence glassy dynamics in surprising ways~\cite{DijkstraABP2013,Berthier2013activeglass,BerthierABP2014,SzamelAOUP2015,FlennerAOUP2016,Flenner2020,Debets2021cage,Keta2022}. Grasping the subtle effect of active motion on the process of vitrification has therefore surfaced as an encouraging line of research and one that unites the comprehensive fields of glassy physics and active matter~\cite{Janssen2019active}. 

For passive materials, one of the few first-principles-based theories that is capable of making reasonable predictions about their glassy dynamics is mode-coupling theory (MCT)~\cite{Janssen2018front,gotze2008complex,Gotze1992,reichman2005mode,Gotze1999} (and its hierarchical extension GMCT~\cite{JanssenPRE2014,Janssen2015a,Janssen2016a,Luo2020_1,Luo2020_2,ciarella2021,Mayer2006,SzamelPRL2003,Wu2005,Luo2021,DebetsGMCT2021}). Requiring only the static structure factor as input, it can predict the full relaxation dynamics via the intermediate scattering function with qualitative, and in few instances even semi-quantitative, agreement. In an attempt to better comprehend active glassy dynamics, and inspired by the previous successes of MCT, several works have recently set out to extend the theory by also including self-propelled particle motion~\cite{Voigtmann2017,SzamelABP2019,SzamelAOUP2016,FengHou2017,Reichert2020modecoupling,Reichert2020tracer,Reichert2021rev,farage2014dynamics,DebetsMCT2022}. This has resulted in a number of so-called active MCTs that qualitatively capture part of the active glassy phenomenology. However, most efforts have so far only considered monodisperse systems, and their mathematical derivations are rather involved,  
which might discourage a broader usage of the theories.

Here, to add to our theoretical understanding of active glassy dynamics, we present an active MCT for mixtures of athermal self-propelled particles that is distinct from,  and conceptually more transparent than, previously introduced versions. We consider an overdamped active system for which we can follow a similar strategy as  typically used for the well-established passive underdamped MCT~\cite{reichman2005mode}. This helps make the derivation more insightful and, since we do not have to specify the time evolution of our active force, allows it to be more general. Interestingly, we show that for a monodisperse system our theory yields the \emph{exact} same equation for the intermediate scattering function as has been previously derived in another (seemingly more involved) mode-coupling theory~\cite{SzamelAOUP2016,SzamelABP2019}. Finally, we test our theory's predictive capabilities for multicomponent systems by comparing it to simulation results obtained for a Kob-Andersen mixture of athermal active Brownian quasi-hard spheres. We find that for all particle types our theory gives qualitatively consistent predictions and thus captures a non-trivial maximum of the dynamics when the persistence length equals the so-called cage length.  


\section*{Theory}
\noindent As our model system we consider an athermal $N$-particle active fluid of volume $V$ which consists of $m$ different species $\alpha$ with component number densities $\rho_{\alpha}=N_{\alpha}/V$. In the overdamped limit, the motion in time $t$ of each individual particle is described via the following equation~\cite{Farage2015,DijkstraABP2013,FengHou2017}
\begin{equation}\label{eom_r}
    \dot{\mv{r}}^{\alpha}_{i} = \zeta_{\alpha}^{-1} \left( \mv{F}^{\alpha}_{i} + \mv{f}^{\alpha}_{i} \right).
\end{equation}
Here, $\mv{r}^{\alpha}_{i}$ denotes the position of the $i$th particle of type $\alpha$, the over-dot the derivative with respect to time, $\zeta_{\alpha}$ the (species-dependent) friction constant, and $\mv{F}^{\alpha}_{i}$ and $\mv{f}^{\alpha}_{i}$ the interaction and self-propulsion force acting on particle $i$ respectively. Normally, one proceeds by also introducing a time-evolution equation for the self-propulsion force. This will turn out not to be necessary for our derivation and we will therefore refrain from doing so to improve the generality of our derivation. Examples of popular model systems include so-called active Ornstein Uhlenbeck (AOUPs)~\cite{SzamelAOUP2016,FlennerAOUP2016,SzamelAOUP2015,BerthierAOUP2017,Flenner2020,FengHou2017} and active Brownian particles (ABPs)~\cite{Voigtmann2017,SzamelABP2019}.

Let us now introduce the velocity of each particle, i.e., $\mv{p}^{\alpha}_{i}=\dot{\mv{r}}^{\alpha}_{i}$, and use it to complement the particle positions as our degrees of freedom (essentially replacing the active forces). Note that for thermal particles this is not feasible since $\mv{p}^{\alpha}_{i}$ would become discontinuous due to the thermal noise term. The joint $N$-particle probability distribution of positions and velocities $P_{N}(\Gamma;t)$ then evolves in time via  
\begin{equation}
\dot{P}_{N}(\Gamma;t)= \Omega P_{N}(\Gamma;t),
\end{equation}
where $\Gamma=(\{\mv{r}_{i}^{\alpha}\},\{\mv{p}_{i}^{\alpha}\} )$ represents the degrees of freedom and $\Omega$ is the Smoluchowski operator which can be inferred from the equations of motion for the positions and velocities (see~\cite{Martin2021} for an example with AOUPs). Now we assume that our system can reach a steady-state characterized by a probability distribution $P^{\mathrm{ss}}_{N}(\Gamma)$ that obeys~\cite{SzamelAOUP2016,SzamelABP2019}
\begin{equation}
\Omega P^{\mathrm{ss}}_{N}(\Gamma)=0.
\end{equation}
Using the steady-state distribution, one can define the time-correlation function $\mv{C}(t)$ of any dynamical vector $\mv{A}(t)$ whose elements $A_{i}(t)=[\mv{A}(t)]_{i}$ are functions of the degrees of freedom as
\begin{equation}
    \mv{C}(t)=\avg{\mv{A}^{*}(0)\mv{A}(t)} = \avg{\mv{A}^{*}e^{\Omega^{\dagger} t}\mv{A}}.
\end{equation}
Here averages $\avg{\hdots}$ are taken with respect to $P^{\mathrm{ss}}_{N}(\Gamma)$, $\mv{A}=\mv{A}(0)$, the asterisk denotes complex conjugation, and the adjoint (or backward) Smoluchowski operator $\Omega^{\dagger}$ is working on everything to its right except the probability distribution. Moreover, by taking the derivative with respect to time and setting it to zero, we also obtain the useful property~$\langle\mv{A}^{*}\dot{\mv{A}}\rangle=\avg{\mv{A}^{*}\Omega^{\dagger}\mv{A}}$.

The strategy of mode-coupling theory is then to choose for the elements of the dynamical vector slow or quasi-conserved quantities of the system. We will focus on the conventional ones usually considered in the derivation of underdamped passive MCT, i.e., the density modes and their respective time-derivatives or current modes~\cite{ciarella2021,reichman2005mode,gotze2008complex}. Note that we thus assume that current modes retain their slow character, which might not necessarily be true in an active matter setting. In particular, we have
\begin{equation}
    \mv{A}=\left[\rho^{1}_{\mv{k}},\hdots,\rho^{m}_{\mv{k}},j^{1}_{\mv{k}} ,\hdots,j^{m}_{\mv{k}}\right]=\left[\gv{\rho}_{\mv{k}}, \mv{j}_{\mv{k}}\right],
\end{equation}
where $\rho^{\alpha}_{\mv{k}}=\sum_{i=1}^{N_{\alpha}}e^{i\mv{k}\cdot\mv{r}_{i}^{\alpha}}/\sqrt{N_{\alpha}}$ and $j^{\alpha}_{\mv{k}}=-i\dot{\rho}^{\alpha}_{\mv{k}}$ are the density and current modes respectively, $N_{\alpha}$ the number of particles of type $\alpha$, and $\mv{k}$ a wave vector which probes the (inverse) length scale of interest. 

Having specified our vector we can then employ the Mori-Zwanzig formalism~\cite{Mori65,Zwanzig60} to develop an equation of motion for its time-correlation function. We define a projection operator onto the subspace spanned by $\mv{A}$ as $\mathcal{P}=\sum_{i,j}|A_{i}\rangle G^{-1}_{ij}\langle A^{*}_{j}|$ (and its complement $\mathcal{Q}=\mathcal{I}-\mathcal{P}$), where we have introduced $\mv{G}=\langle \mv{A}^{*}\mv{A}\rangle$. The superscript $-1$ denotes the inverse matrix of the respective quantity, i.e.\ $G^{-1}_{ij}\equiv [\mathbf{G}^{-1}]_{ij}$. Note that the normalization $G^{-1}_{ij}$ ensures the idempotence of $\mathcal{P}$. Following standard procedure one may find 
\begin{equation}
    \dot{\mv{C}}(t) + \mv{H}\cdot\mv{G}^{-1}\cdot\mv{C}(t)+\int_{0}^{t}dt^{\prime}\  \mv{K}(t-t^{\prime})\cdot\mv{G}^{-1}\cdot\mv{C}(t^{\prime})=0,
\end{equation}
where $\mv{H}=-\avg{\mv{A}^{*}\Omega^{\dagger}\mv{A}}=-\langle\mv{A}^{*}\dot{\mv{A}}\rangle$ denotes the frequency matrix and $\mv{K}(t)=-\avg{\mv{A}^{*}\Omega^{\dagger}\mathcal{Q}e^{\mathcal{Q}\Omega^{\dagger}\mathcal{Q}t}\mathcal{Q}\Omega^{\dagger}\mv{A}}$ the memory function. 

Given the division of our dynamical vector into density and current modes, it is now convenient to segment $\mv{C}(t)$ into four separate $(m \times m)$ matrices as
\begin{equation}
   \mv{C}(t)= \begin{bmatrix}
\mv{C}_{\rho\rho} & \mv{C}_{\rho j}\\
\mv{C}_{j\rho} & \mv{C}_{jj}
\end{bmatrix}.
\end{equation}
Similar to previous work in active MCT we assume that current densities vanish after integrating out the velocities (or active forces)~\cite{SzamelAOUP2016,SzamelABP2019}. In other words, we assume that
\begin{equation}\label{vcurrents}
   \int d\mv{p}_{1}^{1} \hdots d\mv{p}_{N_{\mathrm{m}}}^{m}\  \mv{p}_{i}^{\alpha}P_{N}^{\mathrm{ss}}(\Gamma) = 0.
\end{equation}
Consequently, we have $\langle\rho^{\alpha *}_{\mv{k}}\dot{\rho}^{\beta}_{\mv{k}}\rangle=0$.
If we then use that $\langle\rho^{\alpha *}_{\mv{k}}\rho^{\beta}_{\mv{k}}\rangle=S_{\alpha\beta}(k)$ and $\langle j^{\alpha *}_{\mv{k}}j^{\beta}_{\mv{k}}\rangle=k^{2}\omega_{\alpha\beta}(k)$, it can be shown that  
\begin{equation}
   \mv{G}= \begin{bmatrix}
\mv{S}(k) & \mv{0}\\
\mv{0} & k^{2}\gv{\omega}(k)
\end{bmatrix}, \hspace{0.2cm}
\mv{G}^{-1}= \begin{bmatrix}
\mv{S}^{-1}(k) & \mv{0}\\
\mv{0} & k^{-2}\gv{\omega}^{-1}(k)
\end{bmatrix}.
\end{equation}
Here, we have introduced the partial static structure factor $S_{\alpha\beta}(k)$ and the static velocity-structure correlation function (both of which are symmetric in their species label),
\begin{equation}
\begin{split}
    \omega_{\alpha\beta}(k) = & \frac{\zeta_{\alpha}^{-1}\zeta_{\beta}^{-1}}{\sqrt{N_{\alpha}N_{\beta}}} \uv{k}\cdot \bigg\langle \sum_{i=1}^{N_{\alpha}} \left( \mv{F}^{\alpha}_{i} + \mv{f}^{\alpha}_{i} \right)e^{-i\mv{k}\cdot\mv{r}_{i}^{\alpha}}\times \\
    &\sum_{j=1}^{N_{\beta}}\left( \mv{F}^{\beta}_{j} + \mv{f}^{\beta}_{j} \right)e^{i\mv{k}\cdot\mv{r}_{j}^{\beta}}\bigg\rangle \cdot \uv{k}.
\end{split}
\end{equation}
Interestingly, the latter function is identical to the one introduced in a different active MCT approach~\cite{SzamelAOUP2016} and has subsequently been studied in several numerical works~\cite{FlennerAOUP2016,BerthierAOUP2017,Flenner2020}. It serves to quantify correlations between the velocities of individual particles and represents a distinct nonequilibrium feature. That is, it is a constant in the passive limit and develops significant oscillations upon increasing the persistence of particles.

To find a suitable expression for the frequency matrix we will also assume that the time-correlated version of $\gv{\omega}(k)$ decays exponentially over a timescale equal to the persistence time $\tau_{\mathrm{p}}$ of the active force (assumed to be the same for all species). This corresponds to the fact that the velocities decorrelate over a similar characteristic time as the active forces and implies that we have
\begin{equation}
\langle j^{\alpha *}_{\mv{k}}\dot{j}^{\beta}_{\mv{k}}\rangle \approx -\frac{k^{2}}{\tau_{\mathrm{p}}} \omega_{\alpha\beta}(k)
\end{equation}
and thus
\begin{equation}
   \mv{H}= \begin{bmatrix}
\mv{0} & -i k^{2}\gv{\omega}(k) \\
-i k^{2}\gv{\omega}(k) & k^{2}\gv{\omega}(k)/\tau_{\mathrm{p}}
\end{bmatrix}.
\end{equation}
Note that this approximation likely becomes progressively worse for larger persistence as velocities are probably decorrelating on shorter time scales due to collisions with other particles.

Finally, we realize that
\begin{equation}\label{Rforce_r}
\begin{split}
&\mathcal{Q}\Omega^{\dagger}|\mv{A}\rangle = |\dot{\mv{A}}\rangle + |\mv{A}\rangle \cdot \mv{G}^{-1} \cdot \mv{H} = \Big[ |\mv{0}\rangle,|\dot{\mv{j}}_{\mv{k}}\rangle \\
& \hspace{0.0cm}  
-ik^{2} |\gv{\rho}_{\mv{k}}\rangle\cdot \mv{S}^{-1}(k)\cdot \gv{\omega}(k) + \tau^{-1}_{\mathrm{p}} |\mv{j}_{\mv{k}}\rangle  \Big]\equiv\left[ |\mv{0}\rangle, |\mv{R}_{r}\rangle \right].
\end{split}
\end{equation}
Combining the above derived results and focusing on the lower-left term of $\mv{C}(t)$, whose elements are proportional to the time derivative of the intermediate scattering function $F_{\alpha\beta}(k,t)=\langle\rho^{\alpha *}_{\mv{k}}\rho^{\beta}_{\mv{k}}(t)\rangle$, allows us to write down the following dynamical equation for $F_{\alpha\beta}(k,t)$: 
\begin{equation}\label{eomF}
\begin{split}
    &\ddot{F}_{\alpha\beta}(k,t) + \frac{1}{\tau_{\mathrm{p}}}\dot{F}_{\alpha\beta}(k,t)  +\sum_{\gamma\delta}k^{2}\omega_{\alpha\gamma}(k)S^{-1}_{\gamma\delta}(k)F_{\delta\beta}(k,t) \\
    & + \int_{0}^{t}dt^{\prime}M_{\alpha\gamma}(k,t-t^{\prime})\omega^{-1}_{\gamma\delta}(k)\dot{F}_{\delta\beta}(k,t^{\prime}) = 0.
\end{split}
\end{equation}

The most involved term in this equation is the memory kernel, which is formally written as
\begin{equation}
    M_{\alpha\beta}(k,t)=-\frac{1}{k^{2}}\langle R_{l}^{\alpha} e^{\mathcal{Q}\Omega^{\dagger}\mathcal{Q}t} R_{r}^{\beta} \rangle,
\end{equation}
with $|\mv{R}_{r}\rangle$ as defined in \cref{Rforce_r} and~$\langle \mv{R}_{l}|=\langle \mv{j}_{\mv{k}}^{*}|\Omega^{\dagger} - ik^{2}\gv{\omega}(k)\cdot \mv{S}^{-1}(k) \cdot \langle\gv{\rho}_{\mv{k}}^{*}| + \tau_{\mathrm{p}}^{-1}\langle\mv{j}_{\mv{k}}^{*}|$. To proceed and make analytical progress we need to approximate this term. Therefore, we apply standard techniques from conventional MCT and replace its projected with full dynamics, while also projecting on density doublets~\cite{ciarella2021,DebetsGMCT2021,Nagele1999}. In other words, we have
\begin{equation}
\begin{split}
    &  M_{\alpha\beta}(k,t)\approx -\frac{1}{k^{2}}\langle R_{l}^{\alpha} \mathcal{P}_{2} e^{\Omega^{\dagger} t} \mathcal{P}_{2} R_{r}^{\beta} \rangle \\
    &=-\frac{1}{4k^{2}}\sum_{\mv{q}_{1}\hdots \mv{q}_{4}}\sum_{\mu_{1}\hdots \mu_{4}}\sum_{\nu_{1}\hdots \nu_{4}} \langle R^{\alpha}_{l} \rho^{\mu_{1}}_{\mv{q}_{1}} \rho^{\mu_{2}}_{\mv{q}_{2}} \rangle S^{-1}_{\mu_{1}\nu_{1}}(q_{1})S^{-1}_{\mu_{2}\nu_{2}}(q_{2}) \\
    &\hspace{0.4cm} \times \langle \rho^{\nu_{1} *}_{\mv{q}_{1}} \rho^{\nu_{2} *}_{\mv{q}_{2}} e^{\Omega^{\dagger} t} \rho^{\mu_{3}}_{\mv{q}_{3}} \rho^{\mu_{4}}_{\mv{q}_{4}} \rangle S^{-1}_{\mu_{3}\nu_{3}}(q_{3})S^{-1}_{\mu_{4}\nu_{4}}(q_{4}) \langle \rho^{\nu_{3} *}_{\mv{q}_{3}} \rho^{\nu_{4} *}_{\mv{q}_{4}} R_{r}^{\beta} \rangle
\end{split}
\end{equation}
where we have used the projection operator,
\begin{equation}
    \mathcal{P}_{2}=\frac{1}{2} \sum_{\mv{q}_{1} \mv{q}_{2}}\sum_{\mu_{1} \mu_{2}}\sum_{\nu_{1}\nu_{2}} |\rho^{\mu_{1}}_{\mv{q}_{1}} \rho^{\mu_{2}}_{\mv{q}_{2}} \rangle S^{-1}_{\mu_{1}\nu_{1}}(q_{1})S^{-1}_{\mu_{2}\nu_{2}}(q_{2})\langle \rho^{\nu_{1} *}_{\mv{q}_{1}} \rho^{\nu_{2} *}_{\mv{q}_{2}}|.
\end{equation}

Our aim is now to calculate explicit expressions for the so-called vertices, i.e., $\langle R^{\alpha}_{l} \rho^{\mu_{1}}_{\mv{q}_{1}} \rho^{\mu_{2}}_{\mv{q}_{2}} \rangle$ and $\langle \rho^{\nu_{3} *}_{\mv{q}_{3}} \rho^{\nu_{4} *}_{\mv{q}_{4}} R_{r}^{\beta} \rangle$. For this we first note that due to our assumption of vanishing currents [see \cref{vcurrents}], terms of the form $\langle j^{\alpha *}_{\mv{k}}\rho_{\mv{q}_{1}}^{\beta}\rho_{\mv{q}_{2}}^{\gamma}\rangle$ are equal to zero. Furthermore, we will employ the convolution approximation~\cite{Nagele1999}
\begin{equation}
    \langle \rho^{\alpha *}_{\mv{k}} \rho^{\beta}_{\mv{q}_{1}} \rho^{\gamma}_{\mv{q}_{2}} \rangle \approx \delta_{\mv{k},\mv{q}_{1}+\mv{q}_{2}} \sum_{\lambda}\frac{1}{\sqrt{N_{\lambda}}} S_{\lambda\alpha}(k)S_{\lambda\beta}(q_{1})S_{\lambda\gamma}(q_{2}),
\end{equation}
with $\delta_{ij}$ a Kronecker delta, and make use of the fact that due to time-translational invariance we may rewrite $\langle \dot{j}^{\alpha *}_{\mv{k}}\rho_{\mv{q}_{1}}^{\beta}\rho_{\mv{q}_{2}}^{\gamma}\rangle=-\langle j^{\alpha *}_{\mv{k}}\dot{\rho}_{\mv{q}_{1}}^{\beta}\rho_{\mv{q}_{2}}^{\gamma}\rangle - \langle j^{\alpha *}_{\mv{k}}\rho_{\mv{q}_{1}}^{\beta}\dot{\rho}_{\mv{q}_{2}}^{\gamma}\rangle$. The last expression we require is for terms of the form $\langle \dot{\rho}^{\alpha *}_{\mv{k}}\dot{\rho}_{\mv{q}_{1}}^{\beta}\rho_{\mv{q}_{2}}^{\gamma}\rangle$. In particular, we propose a multi-component extension of a previously introduced convolution approximation for correlation functions involving active particle velocities (see Ref.~\cite{SzamelAOUP2016}). This yields
\begin{equation}\label{conv_active}
\begin{split}
     & \langle \dot{\rho}^{\alpha *}_{\mv{k}}\dot{\rho}_{\mv{q}_{1}}^{\beta}\rho_{\mv{q}_{2}}^{\gamma}\rangle = \frac{\zeta_{\alpha}^{-1}\zeta_{\beta}^{-1}}{\sqrt{N_{\alpha}N_{\beta}}} \mv{k}\cdot \bigg\langle \sum_{i=1}^{N_{\alpha}} \left( \mv{F}^{\alpha}_{i} + \mv{f}^{\alpha}_{i} \right)e^{-i\mv{k}\cdot\mv{r}_{i}^{\alpha}}\times \\
    & \hspace{0.5cm}\sum_{j=1}^{N_{\beta}}\left( \mv{F}^{\beta}_{j} + \mv{f}^{\beta}_{j} \right)e^{i\mv{q}_{1}\cdot\mv{r}_{j}^{\beta}}\sum_{l=1}^{N_{\gamma}}e^{i\mv{q}_{2}\cdot\mv{r}_{l}^{\gamma}}\bigg\rangle \cdot \mv{q}_{1} \approx \\
    & \delta_{\mv{k},\mv{q}_{1}+\mv{q}_{2}} \sum_{\lambda}\frac{1}{\sqrt{N_{\lambda}}} \mv{k}\cdot \mv{q}_{1}  \omega_{\alpha\lambda}(k)\omega^{-1}_{\lambda\lambda}(\infty)\omega_{\lambda\beta}(q_{1})S_{\lambda\gamma}(q_{2}),
\end{split}
\end{equation}
where we have introduced
\begin{equation}
   \omega_{\alpha\beta}(\infty)=\delta_{\alpha\beta}\frac{1}{3\zeta_{\alpha}^{2}N_{\alpha}}\big\langle \sum_{j=1}^{N_{\alpha}}(\mv{F}_{j}^{\alpha}+\mv{f}_{j}^{\alpha})^{2} \big\rangle.
\end{equation}
We emphasize that for a single component system ($m=1$) \cref{conv_active} reduces to the one presented in Ref.~\cite{SzamelAOUP2016}.

Using these results one can then show that the memory kernel simplifies to
\begin{equation}
\begin{split}
    M_{\alpha\beta}(k,t)\approx & \frac{1}{4Vk^{2}}\sum_{\mv{q}\mv{q}^{\prime}}\sum_{\mu\nu}\sum_{\mu^{\prime}\nu^{\prime}} V^{\alpha}_{\mu\nu}(\mv{k},\mv{q}) \langle \rho^{\mu *}_{\mv{q}} \rho^{\nu *}_{\mv{k}-\mv{q}} e^{\Omega^{\dagger} t} \\
    & \times \rho^{\mu^{\prime}}_{\mv{q}^{\prime}} \rho^{\nu^{\prime}}_{\mv{k}-\mv{q}^{\prime}} \rangle V^{\beta}_{\mu^{\prime}\nu^{\prime}}(\mv{k},\mv{q}^{\prime}),
\end{split}
\end{equation}
with the vertices given by
\begin{equation}
\begin{split}
    V^{\alpha}_{\mu\nu}(\mv{k},\mv{q})= & \sum_{\gamma}\frac{1}{\sqrt{\rho_{\gamma}}}\omega_{\alpha\gamma}(k) \big[ \mv{k}\cdot\mv{q}\ \delta_{\gamma\nu}C_{\gamma\mu}(q) \\
    & + \mv{k}\cdot(\mv{k}-\mv{q})\  \delta_{\gamma\mu}C_{\gamma\nu}(\abs{\mv{k}-\mv{q}}) \big],
\end{split}
\end{equation}
which in turn are described by a modified direct correlation function,
\begin{equation}
    C_{\alpha\beta}(q)=\sum_{\gamma\epsilon} \left[ \delta_{\alpha\beta} - \omega^{-1}_{\alpha\gamma}(\infty)\omega_{\gamma\epsilon}(q)S^{-1}_{\epsilon\beta}(q) \right].
\end{equation}
To make our equation self-consistent (and thus solvable) we factorize the four-point density correlation function, i.e.,
\begin{equation}
\begin{split}
    \langle \rho^{\mu *}_{\mv{q}} \rho^{\nu *}_{\mv{k}-\mv{q}} & e^{\Omega^{\dagger} t} \rho^{\mu^{\prime}}_{\mv{q}^{\prime}} \rho^{\nu^{\prime}}_{\mv{k}-\mv{q}^{\prime}} \rangle  \approx  F_{\mu\mu^{\prime}}(q,t) F_{\nu\nu^{\prime}}(\abs{\mv{k}-\mv{q}},t)\ \delta_{\mv{q},\mv{q}^{\prime}} \\
    & + F_{\mu\nu^{\prime}}(q,t)F_{\nu\mu^{\prime}}(\abs{\mv{k}-\mv{q}},t)\ \delta_{\mv{k}-\mv{q},\mv{q}^{\prime}}.
\end{split}
\end{equation}
so that we have
\begin{equation}
\begin{split}
    M_{\alpha\beta}(k,t)\approx & \frac{1}{2k^{2}}\sum_{\mv{q}}\sum_{\mu\nu}\sum_{\mu^{\prime}\nu^{\prime}} V^{\alpha}_{\mu\nu}(\mv{k},\mv{q}) F_{\mu\mu^{\prime}}(q,t) \\
    & \times F_{\nu\nu^{\prime}}(\abs{\mv{k}-\mv{q}},t)  V^{\beta}_{\mu^{\prime}\nu^{\prime}}(\mv{k},\mv{q}),
\end{split}
\end{equation}
and, taking the thermodynamic limit, we finally obtain  
\begin{equation}\label{Mfinal}
\begin{split}
    M_{\alpha\beta}(k,t)\approx & \frac{1}{16\pi^{3}k^{2}}\int d\mv{q}\sum_{\mu\nu}\sum_{\mu^{\prime}\nu^{\prime}} V^{\alpha}_{\mu\nu}(\mv{k},\mv{q}) F_{\mu\mu^{\prime}}(q,t) \\
    & \times F_{\nu\nu^{\prime}}(\abs{\mv{k}-\mv{q}},t)  V^{\beta}_{\mu^{\prime}\nu^{\prime}}(\mv{k},\mv{q}).
\end{split}
\end{equation}
Interestingly, when only considering one particle type ($m=1$), the equations of motion for the intermediate scattering function and specifically the derived memory kernel are identical to the ones for AOUPs and ABPs (neglecting thermal noise) obtained in previous work, which employed a highly different (and seemingly more involved) mode-coupling strategy~\cite{SzamelAOUP2016,SzamelABP2019}. Moreover, in the passive limit where $\tau_{\mathrm{p}}\omega_{\alpha\beta}(k)=D_{\mathrm{t}}\delta_{\alpha\beta}$ (with $D_{\mathrm{t}}$ the translational diffusion coefficient) our equation reduces to the well-known MCT equation derived for mixtures of Brownian particles~\cite{Nagele1999,DebetsGMCT2021}. 

\section*{Methods}
\subsection*{MCT Numerics}
\noindent To self-consistently solve the derived active MCT equations, one needs to complement the theory with a numerical scheme. For this we invoke the rotational symmetry of our system to rewrite the three-dimensional integral over $\mv{q}$ in~\cref{Mfinal} in terms of the bipolar coordinates $q=\abs{\mv{q}}$ and $p=\abs{\mv{k}-\mv{q}}$. The single wavenumber integrals are in turn approximated by a Riemann sum on an equidistant grid $k\sigma = [ 0.2,0.6,\hdots,39.8]$ where $\sigma$ is the unit of length~\cite{Franosch1997}. The integration over time in~\cref{eomF} is evaluated by means of Fuchs' algorithm~\cite{Fuchs1991}. In particular, we evaluate the first $N_{t}/2=16$ points in time via a Taylor expansion with a step size $\Delta t=10^{-6}$, numerically integrate~\cref{Mfinal} for the next $N_{t}/2$ points in time, duplicate the timestep, and repeat this integration process until the long-time limit is reached. 

\subsection*{Simulation Details}
\noindent Given the extension of our theory to multi-component materials, we test its predictions for a model binary active glassformer as a proof of principle. In particular, we choose to study the dynamics of a Kob-Andersen (KA) mixture~\cite{Kob1994} consisting of $N_{\mathrm{A}}=800$ and $N_{\mathrm{B}}=200$ quasi-hard active spheres. The evolution of each particle $i$ is described by~\cref{eom_r}, with the interaction force $\mv{F}_{i}$ derived from a steep repulsive power-law potential $V_{\alpha\beta}(r)= 4\epsilon_{\alpha\beta}\left( \frac{\sigma_{\alpha\beta}}{r}\right)^{36}$~\cite{Weysser2010structural,Lange2009}. The corresponding interaction parameters are given by $\epsilon_{\mathrm{AA}}=1,\  \epsilon_{\mathrm{AB}}=1.5,\  \epsilon_{\mathrm{BB}}=0.5,\  \sigma_{\mathrm{AA}}=1,\ \sigma_{\mathrm{AB}}=0.8,\  \sigma_{\mathrm{BB}}=0.88$, which, combined with setting $\zeta_{\alpha\beta}=1$, allow for glassy behavior by suppressing crystallization~\cite{Kob1994,Michele2004}. For the self-propulsion force $\mv{f}^{\alpha}_{i}$ we use the active Brownian particle (ABP) model~\cite{Voigtmann2017,SzamelABP2019}. That is, the absolute value of the force $f$ remains constant over time, $\mv{f}^{\alpha}_{i}=f\mv{e}^{\alpha}_{i}$, while the orientation $\mv{e}^{\alpha}_{i}$ undergoes random diffusion on the unit sphere, i.e.,
\begin{equation}
    \dot{\mv{e}}^{\alpha}_{i} = \gv{\chi}^{\alpha}_{i} \times \mv{e}^{\alpha}_{i}.
\end{equation}
Here, $\gv{\chi}^{\alpha}_{i}$ represents a Gaussian noise process with zero mean and variance $\avg{\gv{\chi}^{\alpha}_{i}(t)\gv{\chi}^{\beta}_{j}(t^{\prime})}_{\mathrm{noise}}=2D_{\mathrm{r}}\mv{I}\delta_{ij}\delta_{\alpha\beta}\delta(t-t^{\prime})$, with $\mv{I}$ the unit matrix and $D_{\mathrm{r}}$ the rotational diffusion coefficient (taken to be the same for each particle type). 

In the dilute limit when particle-particle interactions are assumed to be absent, each of our particles performs a persistent random walk (PRW). As a result, their mean square displacement (MSD) is given by~\cite{FengHou2017}
\begin{equation}\label{MSDsingle}
    \avg{\delta r^{2}(t)} = 6T_{\mathrm{eff}} \left(\tau_{\mathrm{p}}(e^{-t/\tau_{\mathrm{p}}} - 1) + t \right),
\end{equation}
where we have introduced the persistence time $\tau_{\mathrm{p}}=(2D_{\mathrm{r}})^{-1}$ and an effective temperature $T_{\mathrm{eff}}=f^{2}\tau_{\mathrm{p}}/3$. A closer look at~\cref{MSDsingle} reveals that for $t\ll \tau_{\mathrm{p}}$, the motion is ballistic $\avg{\delta r^{2}(t)}\approx 3T_{\mathrm{eff}}t^{2}/\tau_{\mathrm{p}}$, while diffusive motion, $\avg{\delta r^{2}(t)}\approx 6T_{\mathrm{eff}}t$, is obtained for long times ($t\gg \tau_{\mathrm{p}}$). We can therefore conclude that in the limit $\tau_{\mathrm{p}}\rightarrow 0$ (with $T_{\mathrm{eff}}\sim \mathrm{constant}$), our active system reduces to a Brownian one at a temperature $T=T_\mathrm{eff}$. It is thus convenient to introduce $T_{\mathrm{eff}}$ as our control parameter, which we complement with the persistence length $l_{\mathrm{p}}=f\tau_{\mathrm{p}}$ as a measure for how far we are from the Brownian limit~\cite{Debets2021cage}.

Each individual simulation consists of solving the overdamped equation [\cref{eom_r}] in time with a forward Euler scheme using LAMMPS~\cite{Lammps}. We set the cutoff radius of the repulsive potential at $r_{\mathrm{c}}=2.5\sigma_{\alpha\beta}$ and fix the size of the cubic periodic simulation box to $L=9.41$, such that the number density is $\rho=1.2$. We run the system sufficiently long (typically between $200$ and $1000$ time units) to prevent aging, and afterwards track the particles over time for at least twice the initialization time. All results are presented in reduced units where $\sigma_{\mathrm{AA}}$, $\epsilon_{\mathrm{AA}}$, $\epsilon_{\mathrm{AA}}/k_{\mathrm{B}}$, and $\zeta\sigma^{2}_{\mathrm{AA}}/\epsilon_{\mathrm{AA}}$ represent the units of length, energy, temperature, and time respectively~\cite{Flenner2005}. To correct for the influence of diffusive center-of-mass motion, all particle positions are retrieved relative to the momentary center of mass~\cite{Flenner2005}.

\section*{Results \& Discussion}

\begin{figure}[ht!]
    \centering
    \hspace{-0.3cm}
    \includegraphics [width=0.42\textwidth] 
    {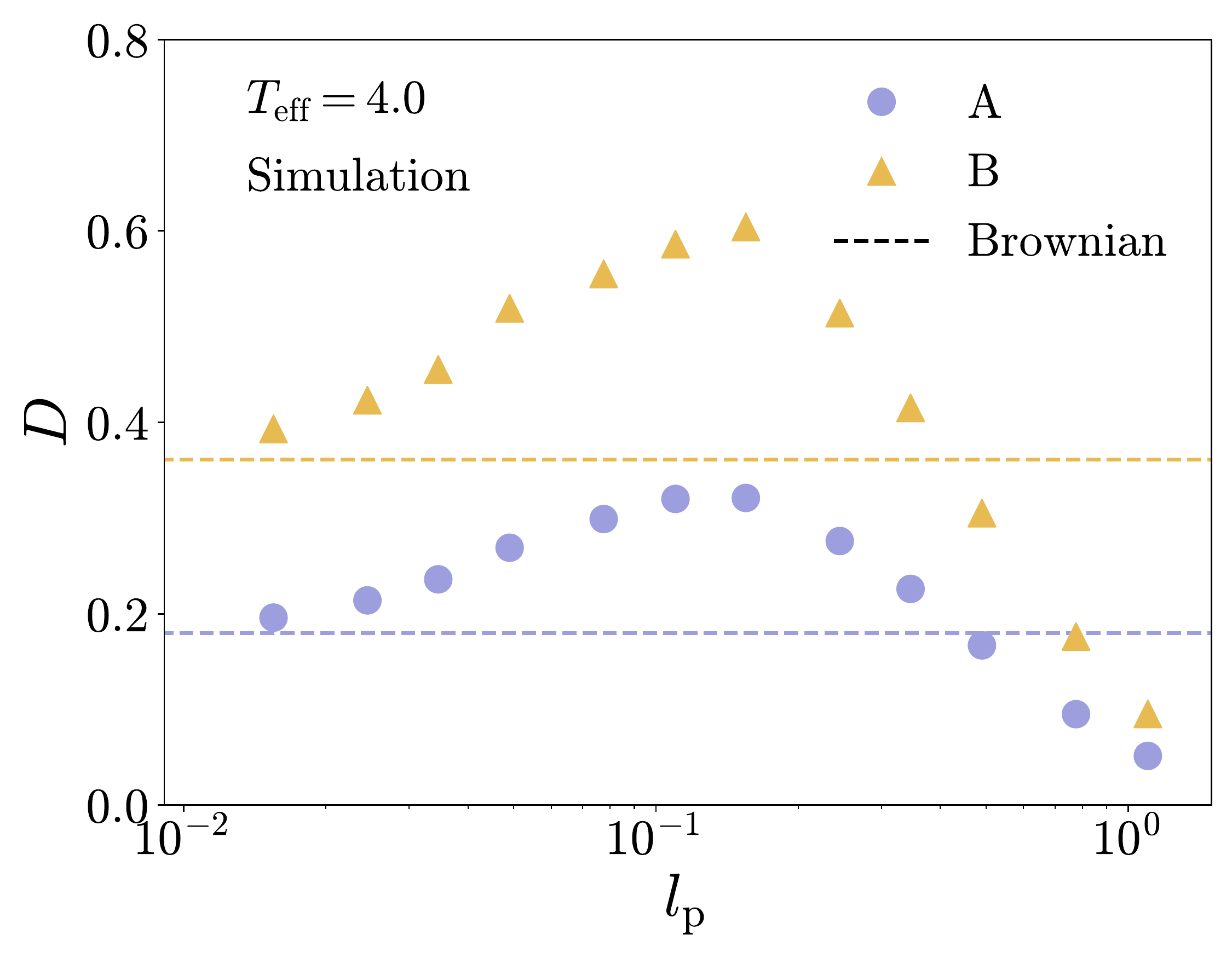} 
    \caption{The long-time diffusion coefficient $D$ of both the A- and B-type particles as a function of the persistence length $l_{\mathrm{p}}$ at a constant effective temperature $T_{\mathrm{eff}}=4.0$. The results are directly obtained from the simulation data and the value retrieved from passive Brownian dynamics simulations ($l_{\mathrm{p}}=0$) is added as a reference (dashed lines).}
    \label{Fig1}
\end{figure}

\noindent In previous work involving the same model glassformer, it has been shown that for a fixed effective temperature $T_{\mathrm{eff}}$, the dynamics exhibits a nonmonotonic dependence on the persistence length $l_{\mathrm{p}}$~\cite{Debets2021cage}. As an initial assessment of the quality of our theory, it is interesting to see whether it is capable of predicting this nontrivial behavior. Before we test our theoretical prediction, however, we first want to verify the nonmonotonic dynamics. For this we have extracted the long-time diffusion coefficient $D=\lim_{t\to\infty}\avg{\delta r^{2}(t)}/6t$ as a function of the persistence length $l_{\mathrm{p}}$ at a fixed value of $T_{\mathrm{eff}}=4.0$. The resulting values are shown in \cref{Fig1} and clearly illustrate nonmonotonic dynamics for both particle species (A and B). Moreover, we find that, consistent with literature~\cite{Debets2021cage}, the maximum of the dynamics corresponds to the point where the persistence length is approximately equal to the cage length, i.e., $l_{\mathrm{p}}\sim 0.1$. 

\begin{figure}[ht!]
    \centering
    \hspace{-0.3cm}
    \includegraphics [width=0.42\textwidth] 
    {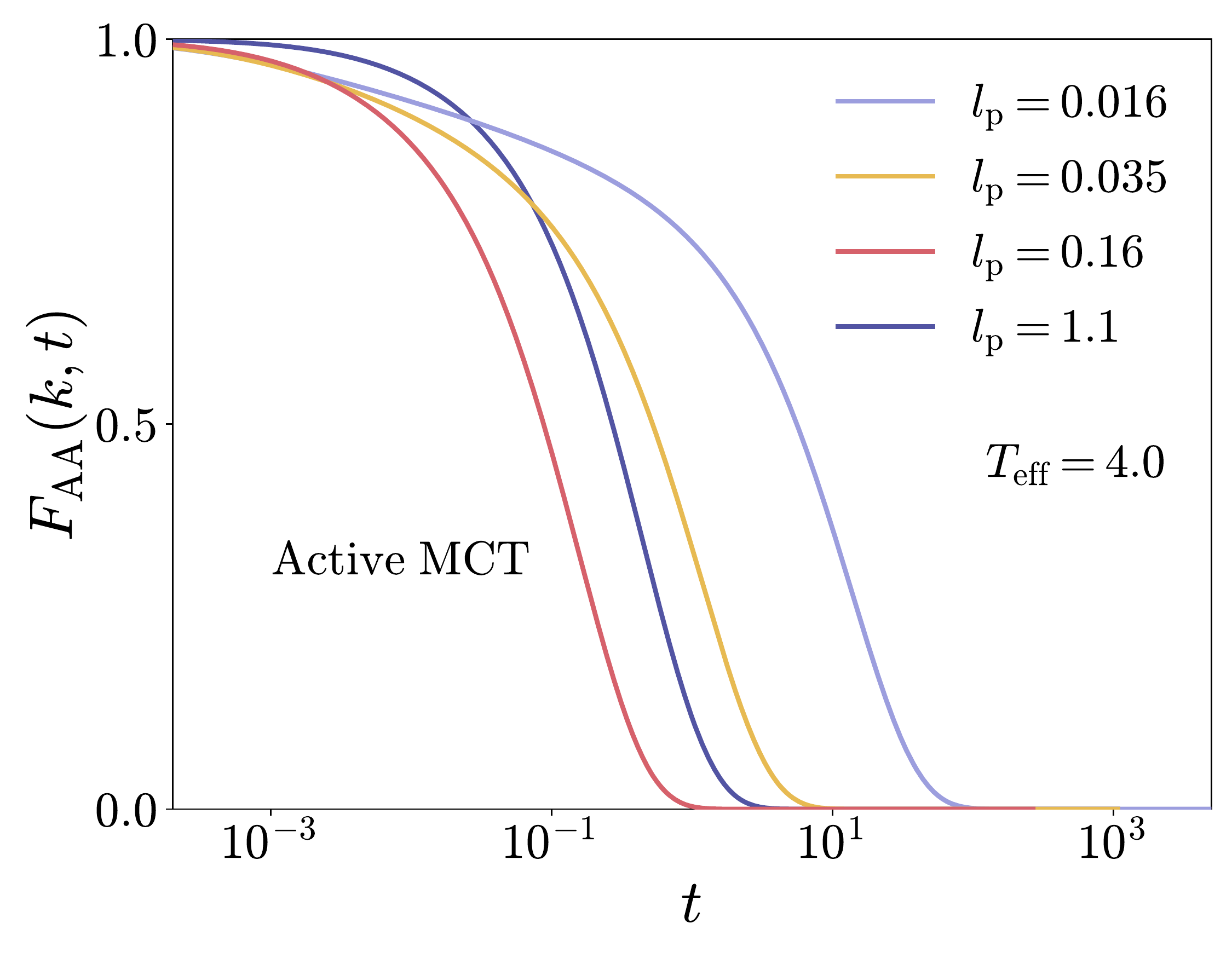} 
    \caption{The intermediate scattering function of the majority A species $F_{AA}(k,t)$ [normalized by the static structure factor $S_{AA}(k)$] as a function of time $t$ for different persistence lengths $l_{\mathrm{p}}$ at a constant effective temperature $T_{\mathrm{eff}}=4.0$. The results are obtained using active MCT for a wavenumber $k$ corresponding to the main peak of the static structure factor.}
    \label{Fig2}
\end{figure}

Having benchmarked our simulation results we now proceed to the active MCT predictions. Based on the retrieved particle trajectories we have calculated the static structure factors $S_{\alpha\beta}(k)$ and velocity-structure correlation functions $\omega_{\alpha\beta}(k)$, which in turn have been rewritten in terms on an equidistant grid via cubic spline. Using these as input for our active MCT, we have calculated the predicted intermediate scattering function $F_{\alpha\beta}(k,t)$. The AA contribution (normalized by the static structure factor) is plotted as a function of time for a subset of persistence lengths in~\cref{Fig2}. Interestingly, this scattering function decays to zero fastest at an intermediate persistence length, and hence the theory seems able to capture the nonmonotonic dependence of the relaxation dynamics.

\begin{figure}[ht!]
    \centering
    \hspace{-0.3cm}
    \includegraphics [width=0.42\textwidth] 
    {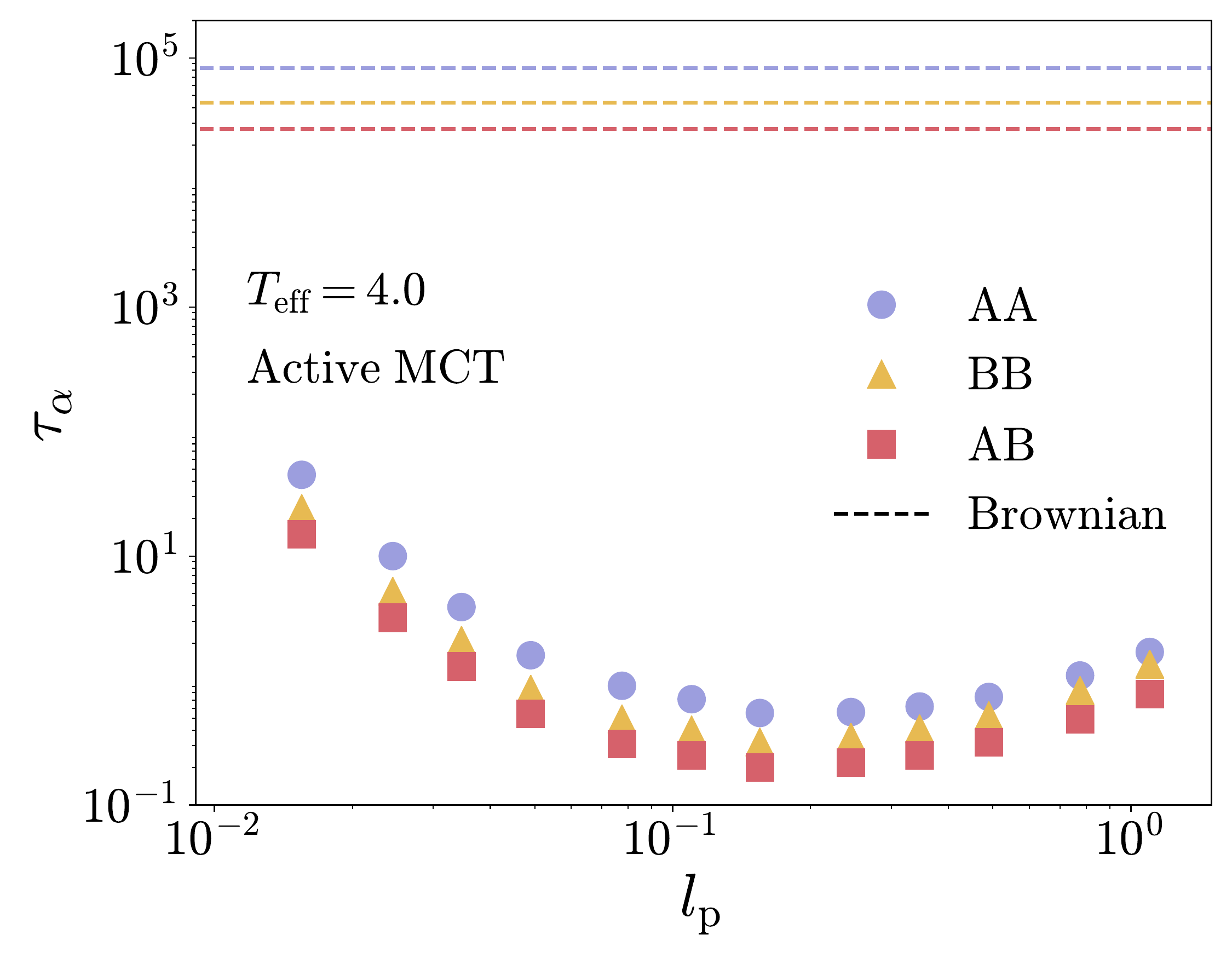} 
    \caption{The alpha relaxation time $\tau_{\alpha}$ as a function of the persistence length $l_{\mathrm{p}}$ obtained from the active MCT intermediate scattering function $F_{\alpha\beta}(k,t)$. Results correspond to an effective temperature $T_{\mathrm{eff}}=4.0$ and are obtained for all particle type combinations. For completeness, the standard MCT prediction based on the structure factor of a passive Brownian system ($l_{\mathrm{p}}=0.0$) is added as a reference (dashed lines).}
    \label{Fig3}
\end{figure}

To quantify the observed behavior in more detail, we have also retrieved the MCT-predicted alpha-relaxation time via $F_{\alpha\beta}(k,\tau_{\alpha})/S_{\alpha\beta}(k)=e^{-1}$ where the wavenumber $k$ corresponds to the main peak of $S_{\alpha\beta}(k)$. The results are presented in~\cref{Fig3} and show clear nonmonotonic and almost identical behavior for all combinations of particle types (either AA, BB, or AB). Note that the fastest relaxation dynamics corresponds to the smallest value of $\tau_{\alpha}$ (in contrast to the largest value of $D$). Moreover, the optimum of the dynamics again coincides with the point where the persistence length is approximately equal to the cage length, i.e., $l_{\mathrm{p}}\sim 0.1$. Our theory is thus capable of accurately depicting the qualitative behavior of the relaxation dynamics upon increasing the persistence of the constituent particles.

\begin{figure}[ht!]
    \centering
    \hspace{-0.3cm}
    \includegraphics [width=0.42\textwidth] 
    {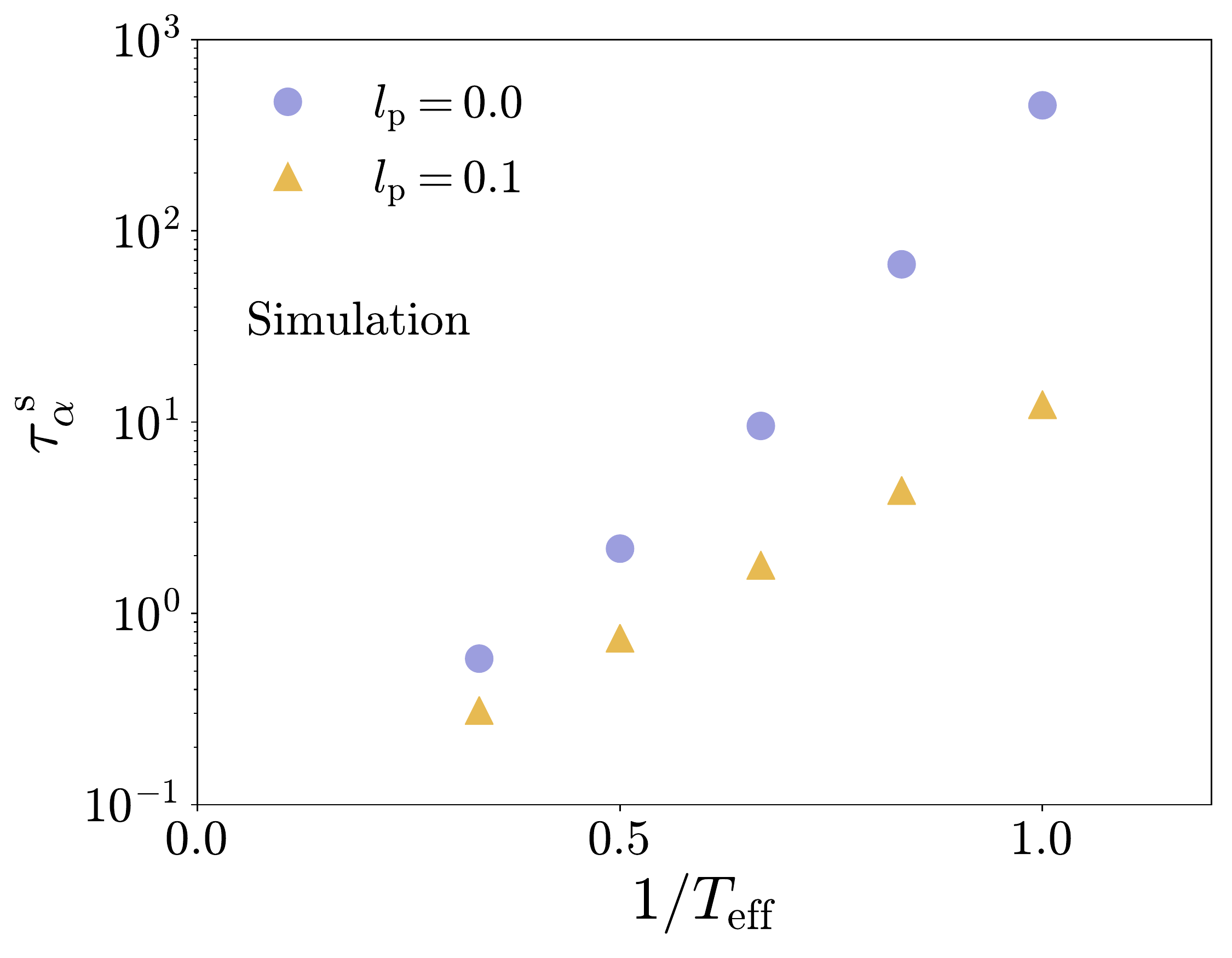} 
    \caption{The alpha relaxation time $\tau_{\alpha}^{\mathrm{s}}$ as a function of the inverse effective temperature for a passive Brownian system ($l_{\mathrm{p}}=0.0$, $T=T_{\mathrm{eff}}$) and an active system with a persistence length on the order of the cage length ($l_{\mathrm{p}}=0.1$). The results are extracted from the self-intermediate scattering of the majority A species $F^{\mathrm{s}}_{\mathrm{A}}(k,t)$ which has been directly calculated from the simulation data.}
    \label{Fig4}
\end{figure}

We finalize our discussion by highlighting two noticeable quantitative features. First we find that in all cases our theory predicts faster relaxation dynamics for the smaller type B particles, which is consistent with our simulation results (see \cref{Fig1}) and is intuitively to be expected~\cite{Flenner2005}. More strikingly, we also observe that our active MCT predicts a dramatic speedup of the dynamics (orders of magnitude decrease of the relaxation time) compared to that obtained from standard MCT for an analogous passive Brownian system at the same effective temperature ($l_{\mathrm{p}}=0.0$, $T=T_{\mathrm{eff}}$). At first glance this might seem surprising, and while it is probably influenced by the assumptions made in the theory, we argue that this behavior is at least partly to be anticipated. To illustrate this we have calculated, based on simulation data, the self-intermediate scattering function, i.e., $F^{\mathrm{s}}_{\alpha}(k,t)=\avg{e^{-i\mv{k}\cdot \mv{r}^{\alpha}_{j}(0)}e^{i\mv{k}\cdot \mv{r}^{\alpha}_{j}(t)}}$, for the majority type A particles and the corresponding alpha relaxation time $\tau^{\mathrm{s}}_{\alpha}$ defined via $F^{\mathrm{s}}_{\alpha}(k,\tau^{\mathrm{s}}_{\alpha})=e^{-1}$. The results for different effective temperatures at a fixed persistence length ($l_{\mathrm{p}}=0.1$, i.e., on the order of the cage length corresponding to the optimum of the dynamics) are shown in~\cref{Fig4}. For comparison, we have also added the values obtained for a passive Brownian system ($l_{\mathrm{p}}=0.0$, $T=T_{\mathrm{eff}}$). We can see that as we lower the (effective) temperature, the relaxation time in both cases increases, but the relative difference between the passive and optimum active dynamics is simultaneously being amplified and can reach differences of several orders of magnitude. Thus, as we approach dynamical arrest by lowering $T_{\mathrm{eff}}$, an optimal active system becomes relatively much more dynamic (i.e., more liquid-like) than its Brownian counterpart. 
Given that our passive MCT predicts an extremely large relaxation time, it is thus to some degree expected that the optimal active dynamics yields a relaxation time that is orders of magnitude smaller.

\section*{Conclusion}
\noindent To conclude, we have presented a fully time-dependent microscopic mode-coupling theory for mixtures of athermal self-propelled particles. The crucial insight for our derivation is that, since we neglect thermal diffusion, the total velocity of each particle is well-behaved; Therefore we can introduce these velocities (instead of the active forces) as our degrees of freedom complementing the particle positions. This then allows us to follow a similar strategy for our overdamped active system as is typically used for passive underdamped MCT~\cite{reichman2005mode}. 
Moreover, it also enables us to leave the time-evolution of the active force unspecified, thereby adding to the generality of the theory. The main result consists of an equation of motion for the (partial) intermediate scattering function, which can be self-consistently solved using the static structure factor and a distinctly non-equilibrium static velocity-structure correlation function as input. Remarkably, for a monodisperse system this equation turns out to be \emph{exactly} equal to one that has been derived in a previous (and possibly more convoluted) active mode-coupling theory~\cite{SzamelAOUP2016,SzamelABP2019}. 


As an initial assessment of the quality of the theory and especially to test its novel extension to multi-component materials, we have used it to predict the dynamics of a Kob-Andersen mixture of athermal active Brownian particles. Such particles exhibit nonmonotonic behavior for increasing particle persistence and thus form a stringent test for the theory. Our theory is indeed able to capture all qualitative features, most notably the location of the optimum of the dynamics when the persistence length and cage length coincide, for each combination of particle types. On a quantitative level active MCT predicts (upon approaching dynamical arrest) a dramatic enhancement of the dynamics (multiple orders of magnitude) compared to that obtained from standard MCT for an analogous passive Brownian system. Though surprising, we show that this effect can in fact be anticipated from simulations.  Given the success of our theoretical framework to give qualitatively consistent results, it would be interesting to see whether the analogy between overdamped athermal active systems and underdamped passive ones can be further exploited to  better understand the intriguing phenomenology of active glassy matter. 

\section*{Acknowledgments}
\noindent We acknowledge the Dutch Research Council (NWO) for financial support through a START-UP grant (V.E.D. and L.M.C.J.).




\bibliographystyle{apsrev4-1}
\bibliography{all}


\newpage






\end{document}